\ifdefined\Nature
\documentclass{nature}
\else
\documentclass[aps,footinbib,letterpaper,superscriptaddress,prl,reprint]{revtex4-1}
\fi

\ifdefined\Nature
\usepackage{cite}
\else
\fi

\newcommand{\expec}[1]{\langle #1\rangle}

\ifdefined\Nature
\newcommand{\subheading}[1]{\noindent \textbf{#1} \\}
\else
\newcommand{\subheading}[1]{\textit{#1.}---}
\fi

\ifdefined\Nature
\newcommand{\labelsubfig}[1]{\textbf{#1,}}
\else
\newcommand{\labelsubfig}[1]{(#1)}
\fi

\ifdefined\Nature
\newcommand{\refsubfig}[1]{#1}
\else
\newcommand{\refsubfig}[1]{(#1)}
\fi

\ifdefined\Nature
\newcommand{\capintro}[1]{\textbf{#1}}
\else
\newcommand{\capintro}[1]{\textbf{#1}}
\fi

\ifdefined\Nature
\newcommand{\scite}[1]{\cite{#1}}
\else
\newcommand{\scite}[1]{~\cite{#1}}
\fi

\ifdefined\Nature
\newcommand{\citein}[1]{~\citen{#1}}
\else
\newcommand{\citein}[1]{~\cite{#1}}
\fi

\ifdefined\Nature
\newcommand{\Ref}[1]{ref.#1}
\else
\newcommand{\Ref}[1]{Ref.#1}
\fi

\ifdefined\Nature
\newcommand{\Refs}[1]{refs#1}
\else
\newcommand{\Refs}[1]{Refs.#1}
\fi

\usepackage{amsmath}
\usepackage{enumitem}
\usepackage{graphicx}

\usepackage[usenames]{color}

\usepackage{hyperref}

\usepackage{braket}

\renewcommand{\v}[1]{\mathbf{#1}}

\newcommand{\lp}{\left ( }
\newcommand{\rp}{\right ) }
\newcommand{\lb}{\left [ }
\newcommand{\rb}{\right ] }

\newcommand{\hc}{\text{h.c.}}

\newcommand{\beq}{\begin{eqnarray*}}
\newcommand{\eeq}{\end{eqnarray*}}

\newcommand{\be}{\begin{eqnarray}}
\newcommand{\ee}{\end{eqnarray}}

\newcommand{\mc}{\mathcal}

\def\lsim{\mathrel{\rlap{\lower4pt\hbox{\hskip1pt$\sim$}}
    \raise1pt\hbox{$<$}}}                

\def\gsim{\mathrel{\rlap{\lower4pt\hbox{\hskip1pt$\sim$}}
    \raise1pt\hbox{$>$}}}                

\ifdefined\Nature
\else
\begin{document} 
\fi

\title{Many-body dynamics of dipolar molecules in an optical lattice}

\ifdefined\Nature{
\begin{document} 
\author{Kaden R.~A. Hazzard$^{1,2}$,  Bryce Gadway$^{1,2}$, Michael Foss-Feig$^{3}$,  Bo Yan$^{1,2}$, Steven A. Moses$^{1,2}$, Jacob P. Covey$^{1,2}$, Norman Y. Yao$^{4}$, Mikhail D. Lukin$^{4}$, Jun Ye$^{1,2}$, Deborah S. Jin$^{1,2}$ \& Ana Maria Rey$^{1,2}$}

\maketitle

\begin{affiliations}
\item JILA, National Institute of Standards and Technology, Boulder, CO 80309, USA
\item Department of Physics, University of Colorado-Boulder, Boulder, CO 80309, USA
\item Joint Quantum Institute, University of Maryland Physics Department, and National Institute of Standards and Technology, Gaithersburg, MD 20899, USA
\item Physics Department, Harvard University, Cambridge, MA 02138, USA
\end{affiliations}
\else
\author{Kaden R.~A. Hazzard} \email{kaden.hazzard@colorado.edu}
\affiliation{JILA, National Institute of Standards and Technology and University of Colorado, Department of Physics, University of Colorado, Boulder, CO 80309-0440, USA}
\author{Bryce Gadway}
\affiliation{JILA, National Institute of Standards and Technology and University of Colorado, Department of Physics, University of Colorado, Boulder, CO 80309-0440, USA}
\author{Michael Foss-Feig}
\affiliation{ Joint Quantum Institute, University of Maryland Physics Department, and National Institute of Standards and Technology, Gaithersburg, MD 20899, USA}
\author{Bo Yan}
\affiliation{JILA, National Institute of Standards and Technology and University of Colorado, Department of Physics, University of Colorado, Boulder, CO 80309-0440, USA}
\author{Steven A. Moses}
\affiliation{JILA, National Institute of Standards and Technology and University of Colorado, Department of Physics, University of Colorado, Boulder, CO 80309-0440, USA}
\author{Jacob P. Covey}
\affiliation{JILA, National Institute of Standards and Technology and University of Colorado, Department of Physics, University of Colorado, Boulder, CO 80309-0440, USA}
\author{Norman Y. Yao}
\affiliation{Physics Department, Harvard University, Cambridge, MA 02138, USA}
\author{Mikhail D. Lukin}
\affiliation{Physics Department, Harvard University, Cambridge, MA 02138, USA}
\author{Jun Ye}
\affiliation{JILA, National Institute of Standards and Technology and University of Colorado, Department of Physics, University of Colorado, Boulder, CO 80309-0440, USA}
\author{Deborah S. Jin}
\affiliation{JILA, National Institute of Standards and Technology and University of Colorado, Department of Physics, University of Colorado, Boulder, CO 80309-0440, USA}
\author{Ana Maria Rey}
\affiliation{JILA, National Institute of Standards and Technology and University of Colorado, Department of Physics, University of Colorado, Boulder, CO 80309-0440, USA}
\fi

\begin{abstract}

Understanding the many-body dynamics of isolated quantum systems is one of the central challenges in modern physics. 
To this end, the direct experimental realization of strongly correlated quantum systems allows one to gain insights into the emergence of complex phenomena. 
Such insights enable the development of theoretical tools that broaden our  understanding.
Here, we theoretically model and experimentally probe with Ramsey spectroscopy the quantum dynamics of disordered, dipolar-interacting, ultracold molecules in a partially filled optical lattice.  We report the capability to control the dipolar interaction strength,
and we demonstrate that the many-body dynamics extends well  beyond a nearest-neighbor or mean-field picture, and cannot be quantitatively described using previously available theoretical tools. 
We develop a novel cluster expansion technique and demonstrate that our theoretical method accurately captures the measured dependence of the spin dynamics on molecule number and on the dipolar interaction strength. 
In the spirit of quantum simulation, this agreement simultaneously benchmarks the new theoretical method and verifies
our microscopic understanding of the experiment. 
Our findings pave the way for numerous applications in quantum information science, metrology, and condensed matter physics.

 \end{abstract}

\ifdefined\Nature
\else

\maketitle

\fi

\begin{figure}[h!]
\setlength{\unitlength}{1.0in}
\includegraphics[width=3.5in,angle=0]{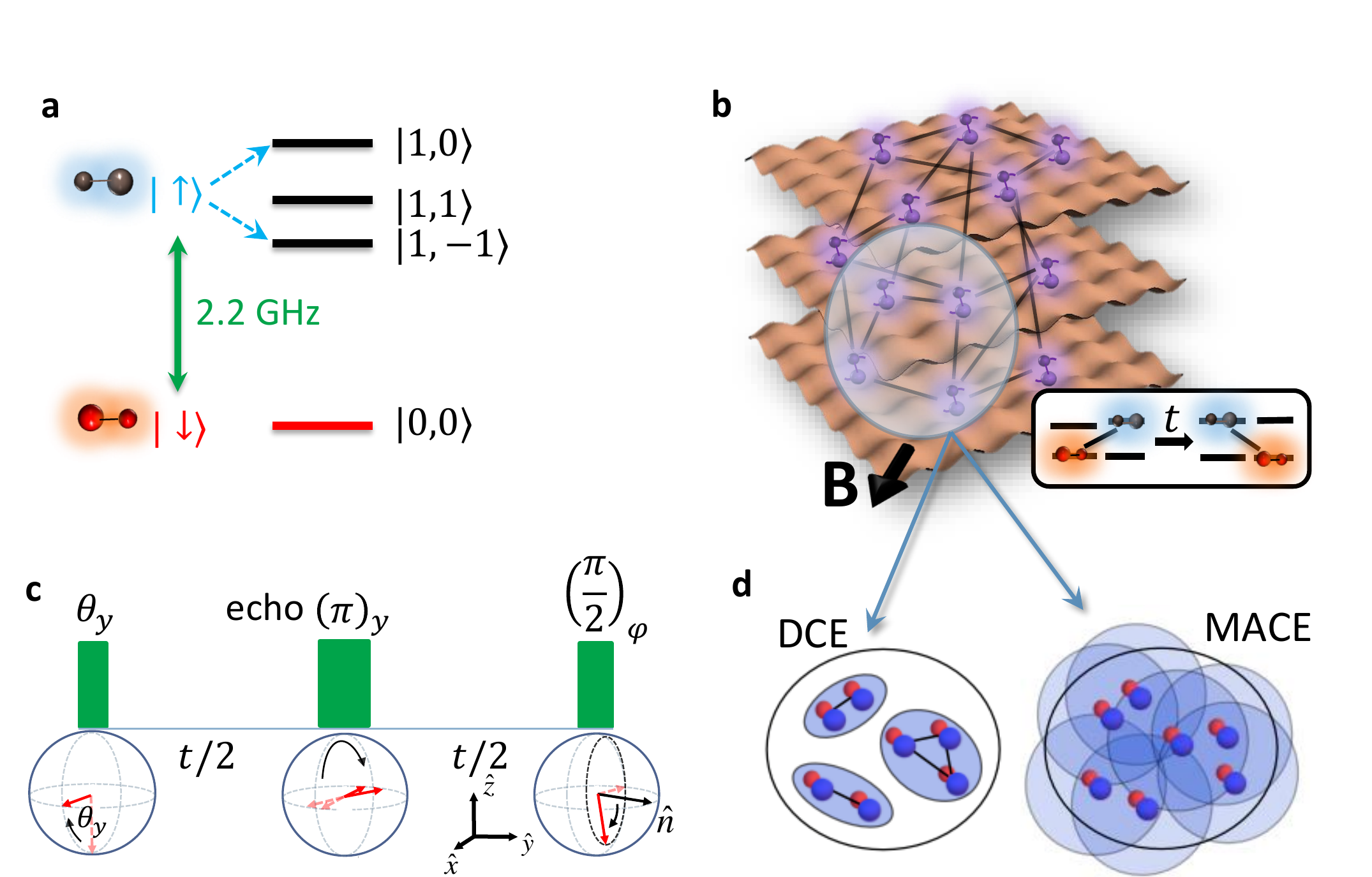}
\caption{\capintro{Probing dipolar spin-exchange interactions of molecules in a lattice with Ramsey spectroscopy.}
 \labelsubfig{a} Two pairs of rotational states in KRb molecules used to realize spin models. The states are labelled $\ket{N, m_N}$, where $N$ is the
total rotational angular momentum  and $m_N$ its projection along the quantization axis set by the magnetic field $\v{B}$. 
\labelsubfig{b}  A dilute gas of KRb molecules pinned in  a deep optical lattice and  experiencing  long-range, anisotropic  dipolar exchange interactions.
\labelsubfig{c}  Ramsey protocol applied to initiate and probe the spin dynamics.
\labelsubfig{d} 
Schematic of the method developed in this paper (MACE), which enables calculations that were intractable with a  disjoint cluster expansion (DCE).
 \label{fig:dynamic-protocol}}
\end{figure}

Advances in trapping and cooling ground-state polar molecules have produced ultracold, nearly degenerate gases\scite{ni_high_2008,ni:dipolar_2010,ospelkaus:quantum-state_2010,
miranda:controlling_2011}, allowed complete control of their hyperfine, rotational, vibrational, and electronic degrees of freedom\scite{ospelkaus_controlling_2010,neyenhuis:anisotropic_2012}, and enabled their preparation in optical lattices\scite{chotia_long-lived_2012}.   Long-range dipolar interactions among molecules facilitate the exploration of fascinating many-body phenomena\scite{baranov:theoretical_2008,
pupillo:condensed_2008,carr:cold_2009,
lemeshko:manipulation_2013,
hazzard:far-from-equilibrium_2013}
and are useful for quantum information processing\scite{demille:quantum_2002}.
By encoding a spin-1/2 degree of freedom in two rotational states,  \Ref\citein{yan:realizing_2013} reported the first observation of dipolar-exchange interactions between molecules in an optical lattice, with signatures such as density-dependence of the spin coherence dynamics as probed by Ramsey spectroscopy (Fig.~\ref{fig:dynamic-protocol}).  Although such dynamics is expected to be governed by a particular spin model\scite{barnett:quantum_2006,gorshkov:tunable_2011}, a quantitative demonstration that the experimental observations  are consistent with this model was lacking.  
A key obstacle was the challenge of modelling three-dimensional, strongly correlated, far-from-equilibrium systems with long-range interactions.

\begin{figure*}
\setlength{\unitlength}{1.0in}
\includegraphics[width=6in,angle=0]{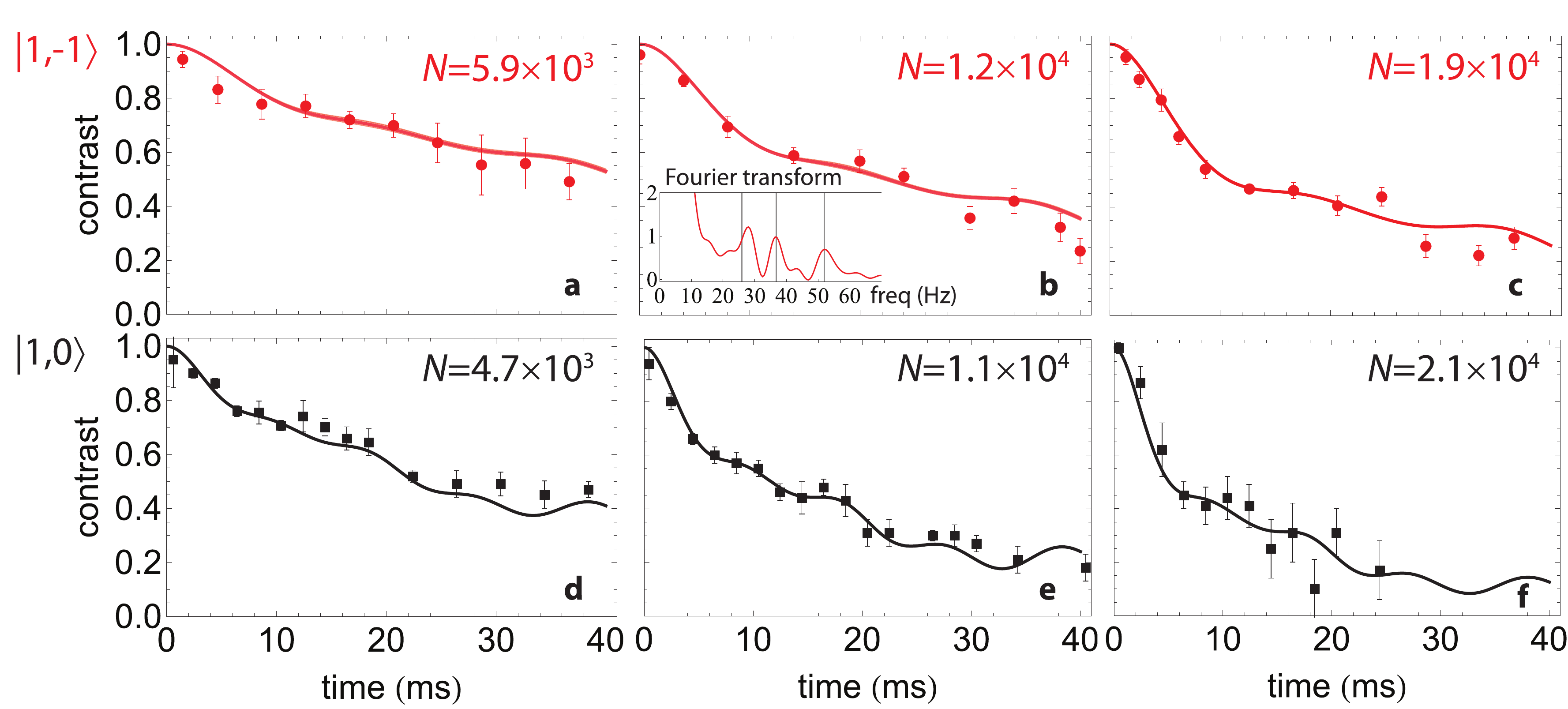}
\caption{ \capintro{Measured contrast dynamics compared with theory.} \labelsubfig{a--c} Contrast versus time for the $\{\ket{\uparrow}=\ket{1,-1},\ket{\downarrow}=\ket{0,0}\}$ rotational state choice, experimentally measured (red symbol) and theoretically calculated (red line) for all spins initially prepared along $\hat x$, for increasing molecule number $N$
(left to right). Inset: Fourier transform of the simulated dynamics for $N=1.2 \times 10^4$ and $\ket{\uparrow}=\ket{1,-1}$.
\labelsubfig{d--f}~Black symbols and black line, same as in (a--c), but choosing instead $\ket{\uparrow}=\ket{1,0}$. 
 \label{fig:theory-v-expt}}
\end{figure*}

In this work, we demonstrate that the  molecules' Ramsey dynamics are quantitatively described by an XY spin model with anisotropic $1/r^3$ interactions and  rule out multiple
alternative models.  This validation  required  parallel experimental and theoretical developments.   Experimentally, we demonstrate that we can tune the dipolar interactions (by a factor of two) by using two different rotational state pairs, and we systematically measure the dynamics under  controlled conditions to quantitatively extract dependences on molecule number.
Theoretically, we develop a new cluster expansion technique, which we term the
``moving-average cluster expansion" (MACE), illustrated
in Fig.~\ref{fig:dynamic-protocol}(d), that is capable of describing the relevant non-equilibrium dynamics of spin models with long range  and anisotropic interactions. MACE converges faster than the prior state-of-the-art disjoint cluster expansion (DCE) and thus enables the description of the spin dynamics of large numbers of molecules.
We benchmark the convergence of MACE in a related (Ising) spin model that is exactly solvable.  However, its convergence in the experimentally relevant XY model is difficult to validate theoretically, because there are no known limits in which this model is amenable to existing numerical methods. 
Ultimately, the convergence is supported by the ability to reproduce the experimental dynamics, making this work an exemplary case of quantum simulation.

\subheading{Polar molecules and long-ranged spin models} Polar molecules  pinned in an optical lattice can realize the long-ranged spin-1/2 model
 \be
H  &=&   \sum_{i\ne j} \frac{V_{ij}}{2}  \lb \frac{J_\perp}{2}   \lp S^+_i S^-_j + \hc\rp + J_z S^z_i S^z_j\rb\!,
\label{eq:XXZ-Ham}
 \ee 
 by encoding the spin in two rotational states\scite{barnett:quantum_2006,gorshkov:tunable_2011,
manmana:topological_2013}.
 (Alternative proposals to implement spin models in polar molecules are discussed in \Refs\citein{micheli_toolbox_2006,buechler:three-body_2007,watanabe:effect_2009,wall:emergent_2009,
 krems:cold_2008,
Schachenmayer:dynamical_2010,perez-rios:external_2010,
trefzger:quantum_2010,kestner:prediction_2011}.)
Here, $S_i^\pm$ and $S_i^z$ are spin-1/2 operators satisfying $ [S_i^z,S_j^\pm]=\pm \delta_{ij} S_i^\pm$, and the sums run over all occupied lattice sites with positions $\v{r}_i$ in units of the lattice spacing $a=532$~nm.  The dipolar interaction couples spins $i$ and $j$  with strength $V_{ij}=(1-3\cos^2\Theta_{ij})/|\v{r}_i-\v{r}_j|^3$, where $\Theta_{ij}$ is the angle between $\v{r}_i-\v{r}_j$ and  the quantization axis, which in our system makes an angle of $45^\circ$ with the $\hat X$ and $\hat Y$ lattice directions [see Fig.~\ref{fig:dynamic-protocol}\refsubfig{b}].  The ``exchange" or ``XY" terms, $S^+_i S^-_j+\hc$, swap the spin states of molecules $i$ and $j$.  These arise from the transition  dipole between the $\ket{\uparrow}$ and $\ket{\downarrow}$ rotational states, and they allow one molecule to flip from up to down while the other flips from down to up, together conserving their combined rotational energy.  The ``direct" or ``Ising" terms, $S^z_i S^z_j$, arise because generally (at finite field) the dipole moments for $\ket{\uparrow}$ and $\ket{\downarrow}$ differ, and thus the parallel and antiparallel configurations of two molecules have different energies.
Note that Eq.~\eqref{eq:XXZ-Ham} does not assume unit filling or even homogeneity;  we will consider the details of the experimental distributions of molecules later.

Figure~\ref{fig:dynamic-protocol} illustrates our experimental system, similar to that described in \Ref\citein{yan:realizing_2013} (Methods and Supplementary Material contain details). We create $N=6,000$ to $23,000$ ground-state fermionic molecules of $^{40}$K$^{87}$Rb in the lattice.  
These molecules have lifetimes exceeding $25\!$~s in the deep 3D optical lattice~\scite{chotia_long-lived_2012}, with the lifetime limited by off-resonant light scattering. In our experiment we work at zero electric field, where the dipole moments and thus $J_z$ vanish, while the resonant exchange coupling $J_\perp$ remains finite. However, 
in order to benchmark the cluster expansions we will also theoretically study the case $J_\perp=0,J_z \ne 0$.

To probe the spin system described by Eq.~\eqref{eq:XXZ-Ham}, we use the Ramsey protocol illustrated in Fig.~\ref{fig:dynamic-protocol}\refsubfig{c}. 
The spins are uniformly rotated to an equal superposition state by a resonant $\pi/2$ microwave pulse about the $\hat{y}$ spin axis, and at a later time $t$ we read out the evolution of the spins by application of a final $\pi/2$ pulse that is phase-shifted by $\varphi$. This final pulse is equivalent to rotation about a vector $\hat{n} = (\sin \varphi, \cos \varphi, 0)$ and measures the quantity $\cos\varphi\sum_i \expec{S_i^x} - \sin\varphi\sum_i \expec{S_i^y}$. 
By varying $\varphi$, we determine the global Ramsey fringe contrast $\mathcal{C}$ defined here as
\be
\mc C &=&  2\big[ \expec{S_i^x}^2+\expec{S_i^y}^2\big]^{1/2} \ .
\ee
We include a $\pi$ spin-echo pulse around $\hat{y}$ at time $t/2$ to remove the dephasing associated exclusively with inhomogeneous light shifts and isolate the effects of spin-spin interactions. 

We  vary the strength of the dipolar interactions  by choosing different pairs of rotational states to realize the spin-1/2 system.  
The exchange coupling $J_\perp$ is determined by the transition matrix dipole element $d_{\uparrow\downarrow}=\braket{\downarrow|d|\uparrow}$,
with $d$ the appropriate spherical component of the dipole operator, via $J_\perp=-d_{\downarrow\uparrow}^2/4\pi \epsilon_0 a^3$ where $\epsilon_0$ is the free space permittivity~\cite{gorshkov:tunable_2011}.
For the states used in \Ref\citein{yan:realizing_2013},
$\ket{\downarrow}= \ket{N=0,m_N=0}$ and $\ket{\uparrow}=\ket{1,-1}$, 
and for our lattice geometry, the transition dipole moment is expected to give an exchange frequency of $|J_\perp / (2h)| \approx 52$~Hz,
where $h$ is Planck's constant. 
For the alternative choice of rotational spin states, $\ket{\downarrow}=\ket{0,0}$ and $\ket{\uparrow}=\ket{1,0}$,
$|J_\perp / (2 h)|$ 
is predicted to be twice as strong ($\sim$100~Hz). This enhancement occurs because the molecules are aligned and oscillate along the quantization axis, as opposed to rotating about it, so that the dipole coupling is not reduced through time-averaging \scite{gorshkov:tunable_2011}. 

Figure~\ref{fig:theory-v-expt}
 compares the observed contrast dynamics as a function of evolution time $t$ for three different molecule numbers, $N \sim 5\times 10^3$, $1\times 10^4$, and $2\times 10^4$, and both pairs of rotational states.
A larger coupling is observed for the $\{\ket{0,0},\ket{1,0}\}$ pair, which is promising for future applications of cold dipolar molecules.  In particular, the faster coherent spin dynamics reduces the effects of technical limitations on the coherence times.
The ability to compare theory and experiment for different exchange couplings is a key feature in our validation of the quantum simulation of the spin model in Eq.~\eqref{eq:XXZ-Ham}.

\subheading{Theory}To simulate the dynamics, we employ a novel cluster expansion technique.  We compare it with a prior cluster expansion, the DCE, which breaks the system of spins into 
non-overlapping (disjoint) clusters, as shown schematically in Fig.~\ref{fig:dynamic-protocol}\refsubfig{d}.  The algorithm to choose these clusters is given in \Refs\citein{witzel:quantum_2005,maze:electron_2008};
roughly, it is designed to  minimize the inter-cluster couplings.  Observables such as the collective spin vector within each cluster  are  exactly computed numerically and then summed over all clusters to obtain the total collective spin dynamics. 

The new approach developed here is to compute $\expec{S_i^\alpha(t)}$ ($\alpha=\{x,y,z\}$) by building an optimal cluster for spin $i$ containing the spins connected to it by the $g$ largest coupling constants $V_{ij}$ -- see Fig.~\ref{fig:dynamic-protocol}\refsubfig{d}.  The dynamics of $\expec{S_i^\alpha(t)}$  is  then computed by exactly solving the dynamics of the entire associated cluster by numerical integration, and
the contrast is obtained by summing these spin expectation values over all $i$.  We term our method the moving average cluster expansion (MACE).
A natural advantage of the MACE method is that it is more robust to artifacts arising from finite cluster sizes; by constructing an optimal cluster for a each spin, MACE reduces surface effects wherein the dynamics of boundary spins in a DCE cluster may not be accurately captured.

We choose a molecule distribution according to our rough expectations of the experimental distribution based on the molecule formation process.  In particular, the molecules are produced only at sites of the lattice initially populated by exactly one Rb and one K atom. Guided by the initial atomic numbers, temperatures, and trapping parameters (see Supplementary Information), we expect a doubly occupied Mott insulator domain of Rb atoms in the centre of the trap, surrounded by a unit-filled ellipsoidal Mott shell where molecules may be formed.  Therefore, we assume a shell of molecules with a filling probability on site $i$ given by $f_i \propto e^{-[(\sqrt{x_i^2+y_i^2+ \alpha^2 z_i^2}-R_c)^2]/(2w^2)}$, with $\alpha = 7$ reflecting the ratio of axial and radial trapping frequencies, and the central shell radius of $R_c=35$.  We determine the  filling fraction by varying the shell width $w$, and set it to match the experimentally observed spin-echo contrast decay time for $N=1.2 \times 10^4$ molecules. Using this procedure, we  find $w=30$. 
Although our calculations use this specific distribution, we find that for a fixed local peak filling of molecules our theoretical results are largely independent of the chosen geometry. 

\subheading{Comparing theory and experiment} 
Figure~\ref{fig:theory-v-expt}  demonstrates that our calculations (solid lines) quantitatively agree  with the measurements for both rotational state choices and 
for a broad range of densities and evolution times.
We emphasize that our results use a single global parameter to reproduce all the experimental data.  This parameter is the ratio of the molecule filling factor $f$ to the molecule number $N$.

\begin{figure*}
\setlength{\unitlength}{1.0in}
\includegraphics[width=6.0in,angle=0]{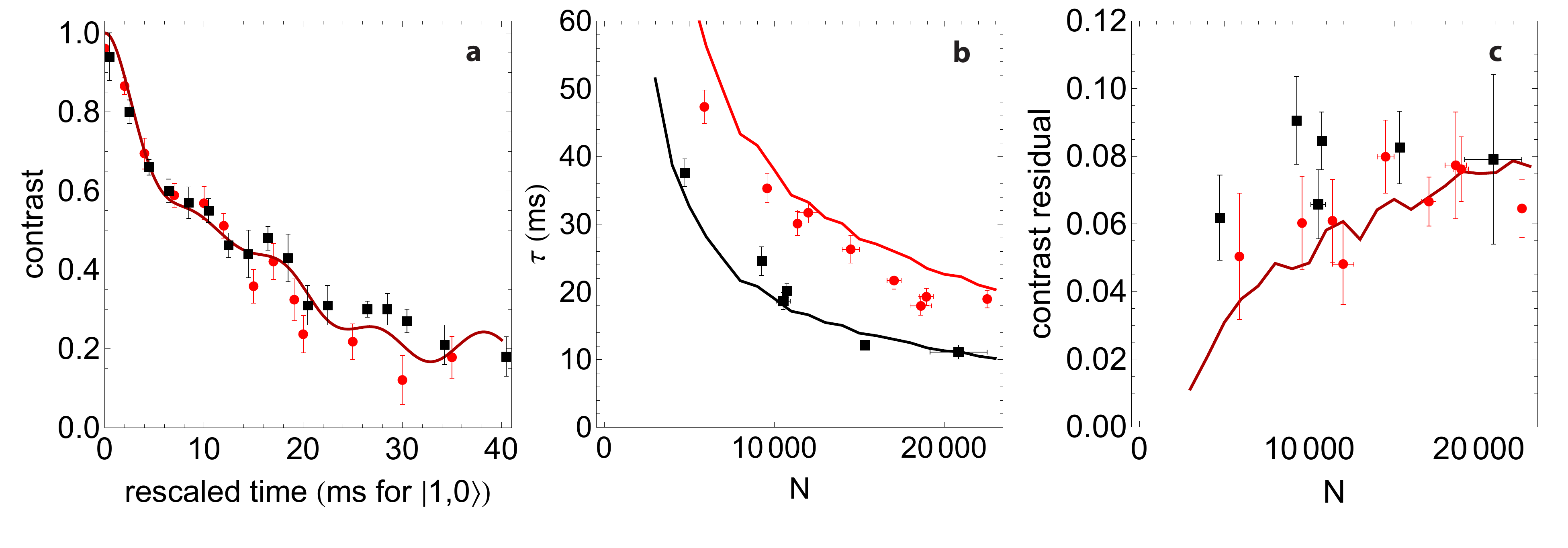}
\caption{ \capintro{Contrast dynamics' dependence on particle number (lattice filling) and rotational state pair choice.} \labelsubfig{a} Rescaled contrast dynamics for the two choices of spin states: the experimental times for the $\ket{\uparrow}=\ket{1,-1}$ data (red circles) are rescaled by a factor of $J_{\ket{1,-1}}/J_{\ket{1,0}}\approx 1/2$ with respect to the $\ket{\uparrow}=\ket{1,0}$ data (black squares), to show that the choice of rotational states only rescales the interaction timescale. Theoretically calculated dynamics (solid lines) is shown for $N=12,000$.
The two measurements have slightly different particle numbers: $N=1.1\times 10^4$ and $N=1.2\times10^4$ for $\ket{\uparrow}=\ket{1,0}$ and $\ket{\uparrow}=\ket{1,-1}$, respectively.
\labelsubfig{b} Spin coherence time $\tau$ versus molecule number $N$, as determined by fitting the contrast data to a simple exponential decay, of the form $\mc{C}(t) = \exp\lp -t/\tau\rp$.
\labelsubfig{c} Root-mean-square residuals of the data from the exponential fit as a function of the molecule number $N$, quantifying the magnitude of all oscillations. Coherence times and residuals are shown for both sets of spin states, with theory predictions as well, with the same colours and symbols as in panel~a.
 \label{fig:theory-v-expt2}}
\end{figure*}

We note a few interesting trends that emerge in both the theory and experimental data, then describe in the following paragraphs each of these trends in more detail. First, as in \Ref\citein{yan:realizing_2013}, we observe clear oscillations of the contrast. These oscillations are roughly independent of the molecule number $N$, but the frequency is found to be larger for the $\ket{1,0}$ data, consistent with the enhanced spin-exchange coupling. Second, we observe that the spin coherence time decreases with an increase in lattice filling. This is a clear signature of spin-spin interactions. In comparing panels (a--c) with (d--f) of Fig.~\ref{fig:theory-v-expt}, this coherence time is also seen to be shorter for the spin states with larger spin-exchange coupling ($\ket{\uparrow} \equiv \ket{1,0}$). Lastly, we find a trend of increasing oscillation amplitude for increasing molecule number.

The contrast oscillations arise from dipole-dipole interactions between the molecules.
Clear oscillations at frequency $|J_\perp/(2h)|$ are visible for all measurements and calculations. Additionally, the Fourier transform of the theoretical contrast  (Fig.~\ref{fig:theory-v-expt}, inset) clearly shows multiple oscillation frequencies on a broad, structured background.  These frequencies are roughly determined by the size of the strongest couplings.
 The most prominent contributions apparent in the  time evolution data appear at the frequencies  $\nu/\sqrt{2}$, and $\nu/2$ (next nearest neighbor coupling strengths), where $\nu = 104$~Hz for  $\ket{\uparrow}=\ket{1,0}$ data and 52~Hz for $\ket{\uparrow}=\ket{1,-1}$. In the Supplementary Materials, we have analyzed the experimental data by fitting to functional forms that oscillate at either a single frequency $\nu$ or at three frequencies. Based on a global analysis of fifteen data sets, we find clear evidence that a multi-frequency fit better captures the observed dynamics. Moreover, the analysis suggests a fundamental frequency of $\nu \sim 108$~Hz for the $\ket{\uparrow}= \ket{1,0}$ data (reduced by half for the $\ket{1,-1}$), in excellent agreement with the expected value.
 The influence of the choice of rotational states is further illustrated in Fig.~\ref{fig:theory-v-expt2}\refsubfig{a}. This plot overlays contrast dynamics for  the two pairs of rotational states, with the times of the $\ket{\uparrow}=\ket{1,-1}$ data rescaled by a factor of 1/2 to account for the different dipolar interaction strength. The collapse of the two datasets 
 highlights that all of the observed dynamics arise from the dipolar interactions.

To investigate the coherence time $\tau$'s dependence on the particle number $N$, 
we experimentally reduce the number of molecules using single-particle loss due to off-resonant light scattering.
This leaves the distribution of molecules invariant by reducing the density of particles uniformly (see Methods).
We extract coherence times by fitting the contrast dynamics to  $\exp\lp-t/\tau\rp$ for both the theory and experiment; the results are shown in Fig.~\ref{fig:theory-v-expt2}\refsubfig{b}. Whereas the oscillations come mainly from the largest frequencies (smallest spacings) between molecules, the decoherence
arises from interactions of molecules at a variety of spacings.   We expect $\tau \propto 1/N$, which is a characteristic signature of interactions. 
The scaling arises because, as the lattice filling $f\propto N$ increases, the mean distance between molecules decreases as ${\bar R}\sim f^{-1/3}$, leading to an average dipolar interaction that scales as $1/{\bar R}^3 \sim f$. We indeed see an approximate scaling $\tau \propto 1/N$, as well as a factor of  2 reduction in coherence time for the $\ket{1,0}$ data compared to the $\ket{1,-1}$ data. We note that the theory here uses a peak filling of $f= 8\%$ for a molecule number of $N = 2 \times 10^4$, which is within a factor of two of estimates based on loss measurements ($\sim  \! 9\%$ for $N\sim1\times 10^4$) and direct imaging\scite{yan:realizing_2013,zhu:direct_2013}.

Finally, the  oscillation amplitude increases with $N$  for the experimental fillings studied since the probability of a molecule having an occupied nearest neighbour site increases with $N$. 
To characterize the amplitude of the oscillations, we plot in Fig.~\ref{fig:theory-v-expt2}\refsubfig{c} the root-mean-square (rms) residuals of the data from the exponential fit. 
We find
that for increasing molecule number $N$ there is a systematic increase of the residuals, due to oscillatory dynamics absent in the simple exponential fit. 

\begin{figure}
\setlength{\unitlength}{1.0in}
\includegraphics[width=3.4in,angle=0]{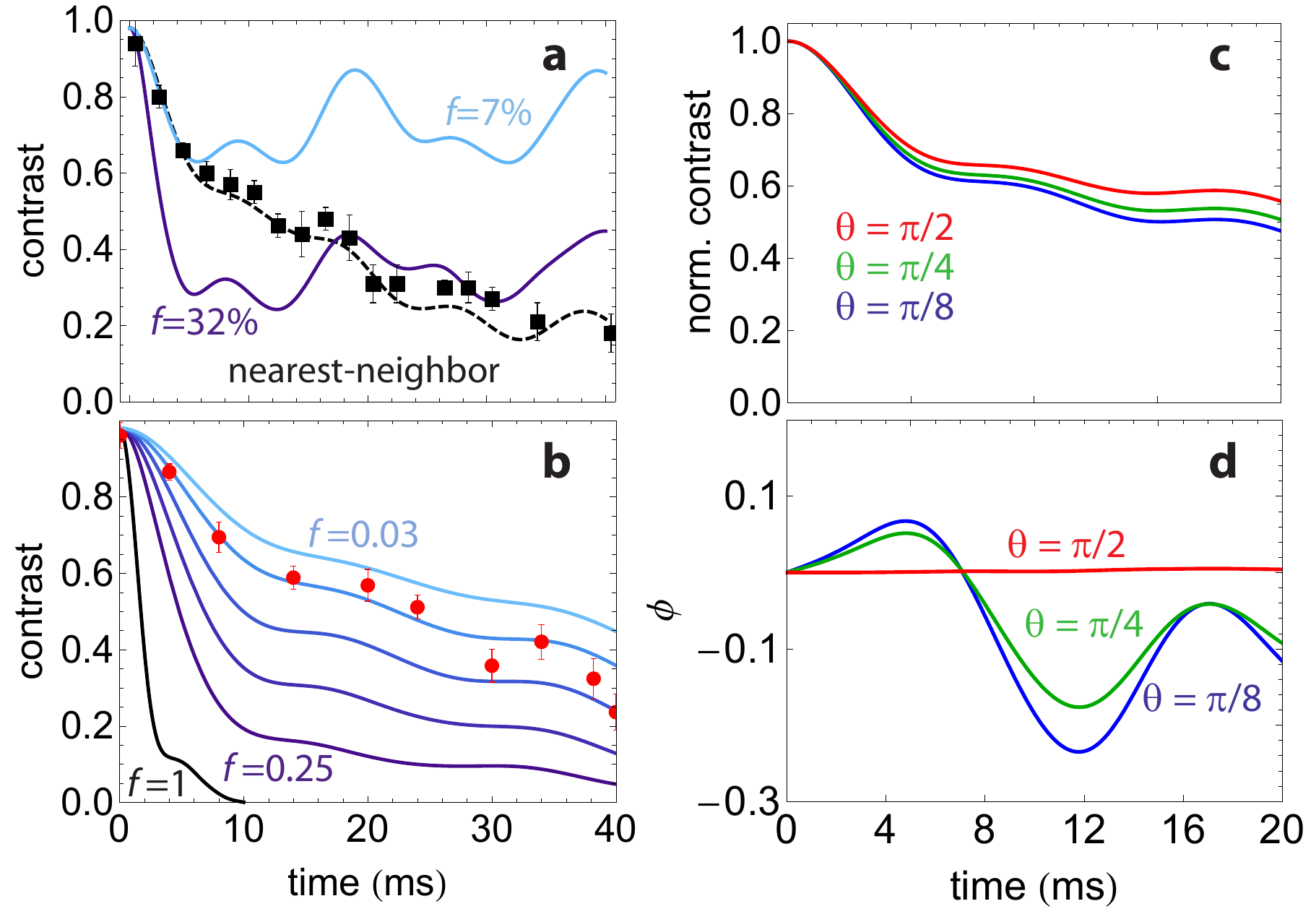}
\caption{\capintro{Predictions for other initial spin directions and higher lattice fillings.} \labelsubfig{a} Nearest-neighbor interaction contrast dynamics for two fillings compared with our $N=1.2\times 10^4$, $\ket{\uparrow}=\ket{1,0}$ measurement. \labelsubfig{b} Contrast dynamics for various fillings from $f=0.025$ to $f=1$, compared with $N=1.1\times 10^4$ molecule measurements using $\ket{\uparrow}=\ket{1,-1}$ (red squares). 
\labelsubfig{c} Near-collapse of scaled contrast versus time [${\mc C}(t)/{\mc C}(t=0)$] for initial spin angles $\theta=\{\pi/2,\pi/4,\pi/8\}$ (top to bottom)  for different $\theta$.  \labelsubfig{d} Bloch vector phase $\phi$ dynamics for $\theta\ne\pi/2$, which is more complex than simple precession.  Panels (c) and (d) are for $\ket{\uparrow}=\ket{1,0}$.
 \label{fig:predictions}}
\end{figure}

\subheading{Discussion} 
Given the accuracy of our theoretical cluster expansion (see Methods and Supplementary Information), our measurements can rule out multiple alternatives to the spin-exchange model Eq.~\eqref{eq:XXZ-Ham} for describing the experimental observations. For example, the experimentally measured dynamics is inconsistent with
the Ising model, where the contrast oscillation amplitudes are significantly smaller than those observed experimentally (see Methods and Fig.~\ref{fig:theory-v-theory} of the Supplementary Information  
for a demonstration.)
As another example, Fig.~\ref{fig:predictions}\refsubfig{a} shows  that  
the spin-exchange model with interactions truncated to the nearest neighbours in the lattice also fails to reproduce our measurements even qualitatively.


We can now make predictions for future experiments, for example
the effects of increasing the filling and varying the initial spin preparation.
Figure~\ref{fig:predictions}\refsubfig{b} shows the effects of increasing the filling towards unity. The contrast decay becomes even more rapid. The contrast oscillation amplitude initially increases monotonically with increasing filling (see Fig.~\ref{fig:predictions}) and then saturates around fillings of 25\%, and finally decreases.
At even higher fillings, the dynamics becomes overdamped and the contrast oscillation amplitude decreases, similar to the one-dimensional behavior studied in \Ref\citein{hazzard:far-from-equilibrium_2013}.

We find an interesting feature when we examine the dependence on the initial spin tipping angle.
For arbitrary tipping angles $\theta\neq\pi/2$, the rescaled contrast ${\mc C(t)}/\sin\theta$ is largely independent of $\theta$, as shown in Fig.~\ref{fig:predictions}\refsubfig{c}.
The contrast decay arises from two effects: dephasing of precession angles between spins, since the mean-field precession rate differs spin to spin (this only plays a role for $\theta \ne \pi/2$), and contrast decay  due to the buildup of correlations.  If $\sum_{ij} V_{ij}=0$, as it nearly does for the anisotropic dipolar interaction, one can show using the analytic formulas  for short-time dynamics in \Ref\citein{hazzard:far-from-equilibrium_2013} that  these two terms balance in such a way that $\mc C(t)/\sin(\theta)$ is independent of $\theta$. It is very interesting that the collapse of the curves in Fig.~\ref{fig:predictions}\refsubfig{c} holds to  longer times. Our theory also predicts that the phase $\phi= \arcsin\lb \expec{\sum_i S_i^y}/\expec{\sum_i S_i^x }\rb$  diffuses randomly and remains small, rather than undergoing a mean-field-like precession, as shown in Fig.~\ref{fig:predictions}\refsubfig{d}.

\subheading{Summary and conclusions}   We have experimentally measured the Ramsey contrast dynamics in ultracold KRb molecules, varying both the molecule density and the strength of the dipolar interactions.  We have developed a theoretical method  that allows us to efficiently and accurately compute the spin coherence evolution of macroscopic numbers of molecules.
The theory quantitatively reproduces the experimental data, and our comparison allows us to determine a typical experimental lattice filling fraction between $5$-$10\%$, which is roughly consistent with that obtained by independent measurements\scite{zhu:direct_2013}.  One question  for future investigation is the role of inhomogeneous light shifts, which add a spatially varying $\sum_i \Delta_i S^z_i$ term in Eq.~\eqref{eq:XXZ-Ham} and can suppress exchange interactions. If we use values of $\Delta_i$ expected from AC polarizabilities~\cite{neyenhuis:anisotropic_2012}, our data and theory begin to disagree at times longer than about $20$~ms.  
This could arise from mis-estimates of the polarizabilities, spatial location of the molecules, or from unanticipated technical sources of decoherence. 

The results presented here exemplify how a clean ultracold system can guide the development of theories applicable to the numerous quantum systems governed by similar Hamiltonians.  For example,  MACE can help understand recent experiments using room-temperature polycrystalline molecular solids\scite{alvarez:nmr_2010},
or atomic optical lattice clocks~\scite{chang:controlling_2004,martin:quantum_2013}.
We
expect the MACE method  to  have applications to numerous
other systems described by long-range spin models~\scite{lahaye:strong_chromium_2007,lahaye:physics_2009,
saffman:quantum_2010,lanyon:universal_ion_2011,
nipper:atomic_2012,lu:quantum_2012,paz:resonant_2012,
britton:engineered_2012,islam:emergence_2013}, such as magnetic atoms, molecules, Rydberg atoms,
trapped ions, solid-state defects, as well
as exciton transport in molecules, relevant for chemistry
and biology\scite{
xiang:tunable_2012,
saikin:photononics_bio_2013}.

In the future it will be fascinating to examine the development of correlations more directly, both theoretically and experimentally, and thereby explore transport/thermalization or lack thereof, e.g. glassiness and many-body localization\scite{basko:many-body_2006,pal:many-body_2010,
kwasigroch:bose-einstein_2013,yao:many-body_2013}.

\ifdefined\Nature
\else
\textit{Acknowledgements.}---We thank Bihui Zhu, Salvatore Manmana, Tilman Pfau, John Bollinger, Joe Britton, Brian Sawyer, Johannes Schachenmayer, Jake Taylor, and Alexey Gorshkov for discussions. The authors acknowledge funding from JILA-NSF-PFC 1125844, NSF-PIF, ARO, ARO-DARPA-OLE, and AFOSR.  S.A.M. and J.P.C. acknowledge
funding from NDSEG.  K.R.A.H., B.G., and M.F.-F. thank the National Research Council postdoctoral fellowship programme for support.  K.R.A.H. and A.M.R. acknowledge the KITP.
\fi

\ifdefined\Nature
\begin{methods}
\else 
\vspace{0.1in}
{\hspace{1.15in} \textbf{Methods.}}  
\fi

\subsection{Experimental production of ultracold KRb in an optical lattice}

To produce the molecules in a lattice, we follow the experimental procedure detailed in \Ref\citein{yan:realizing_2013}.  Briefly summarizing, we create between $N=6,000$ to $23,000$ ground-state molecules in a lattice of spacing $a=532$~nm.  We start from $^{40}$K atoms initially at $T/T_F\approx0.5$ and $^{87}$Rb atoms initially at $T/T_c\approx0.5$ ($T_F$ and $T_c$ are the Fermi and Bose-Einstein condensate transition temperatures, respectively), confined in traps with 25~Hz radial and 185~Hz axial frequencies for Rb. We  create Feshbach molecules by ramping the magnetic field across a broad interspecies Feshbach resonance at $\sim$546~G, and then transfer these to ground-state molecules by a stimulated Raman adiabatic passage (STIRAP) process. To realize spin systems, we freeze out the tunnelling by working in a deep 40~$E_r$ lattice, where $E_r=\hbar^2 k^2/(2m)$ is the photon recoil energy with momentum $k=\pi/a$, where $m$ is the molecule mass. We image by reversing the STIRAP process and then taking  absorption images of the K atoms after release from the lattice.

In order to change the filling of molecules uniformly, leaving their distribution unaltered, we employ single particle loss from light scattering.  Specifically,
we hold the molecules in the deep lattice with two additional running beams that are homogeneous over the scale of the cloud.  The time for which these are applied is chosen to achieve the desired density.

\subsection{Convergence of theory}
We demonstrate the convergence of our theoretical method (MACE), validating our theoretical calculations.
We also discuss the  advantages of the MACE over prior methods (DCE). 
We demonstrate convergence for the spin-exchange dynamics, but to further verify the accuracy and convergence to the correct solution, we also consider the 
$J_\perp=0$ case, which admits an exact analytic treatment\scite{emch:non-markovian_1966,kastner:diverging_2011,hazzard:far-from-equilibrium_2013,foss-feig:nonequilibrium_2013}.
For example, for an initial tipping angle of the spins relative to $-{\hat z}$, $\theta=\pi/2$, the exact Ramsey contrast is
${\mc C}_{\text{Ising}} = \left | \sum_i \lb \prod_{j\ne i} \cos\lp J_z V_{ij} t/2\rp \rb \right | $ ($\hbar=1$).

Figure~\ref{fig:theory-v-theory} in the Supplementary Information illustrates the convergence properties of the DCE and MACE.  Dashed (solid) lines
show the contrast versus time for the DCE~(MACE) with cluster sizes $g=4,6,8,10$, in both cases for $N=5000$ spins in a  ellipsoidal shell with probability distribution
$f_i  \propto e^{-[(\sqrt{x_i^2+y_i^2+ \alpha'^2 z_i^2}-R_c')^2]/(2w'^2)}$ with
$R_c'=25$, $\alpha'=3$, and $w'=25$. 
  Figure~\ref{fig:theory-v-theory}\refsubfig{a} shows the computed XY dynamics ($J_z=0$), while the panel~\refsubfig{b} shows Ising dynamics ($J_\perp=0$) compared with the exact Ising solution (open circles).
In both cases, the MACE is quantitatively accurate even for small clusters $g\sim 6$. In contrast, although the oscillations  at shorter times are well-captured by the DCE, the long time behaviour is poorly converged, even for $g=10$.  For the Ising case, 
despite the apparent convergence of the DCE we find that a large discrepancy remains even for enormous cluster sizes of $g\sim 500$.  These statements are complemented in the Supplementary Information by a quantitative analysis of the root-mean-square error, both between successive cluster sizes and for the Ising case between the approximate and exact solution. 
The reason the DCE fails is that the dynamics of spins at cluster boundaries is captured poorly, while in the MACE each cluster is optimized around the molecule contributing to the spin expectation, avoiding these problems. Although for sufficiently large clusters these boundary contributions are negligible and both methods formally converge to the same result, the cluster sizes required by the DCE to reach this limit are far larger than numerically possible. 
We also note that the MACE is trivially parallelizable: each spin can be considered independently.  In contrast, the creation of clusters in  the DCE is less trivial to parallelize and can hinder scaling to large numbers of spins.

\ifdefined\Nature
\end{methods}
\else	
\fi


\ifdefined\Nature
\begin{addendum}
\item[Supplementary Information\!\!\!\!] is available in the online version of the paper.
\item[Acknowledgements\!\!\!\!] We thank Bihui Zhu, Salvatore Manmana, Tilman Pfau, John Bollinger, Joe Britton, Brian Sawyer, Johannes Schachenmayer, Jake Taylor, and Alexey Gorshkov for discussions. The authors acknowledge funding from JILA-NSF-PFC 1125844, NSF-PIF, ARO, ARO-DARPA-OLE, and AFOSR. K.R.A.H., B.G., and M.F.-F. thank the National Research Council postdoctoral fellowship programme for support. S.A.M. and J.P.C. acknowledge
funding from NDSEG.  K.R.A.H. and A.M.R. acknowledge the KITP.
\item[Author Contributions\!\!\!\!] The experimental work was carried out by  B.G., B.Y., S.A.M., J.P.C.,  J.Y., and D.S.J. Theoretical modelling and calculations were done by K.R.A.H., M.F.-F. and A.M.R.  Analysis and comparison of experimental data and theoretical calculations was carried out by K.R.A.H., B.G., M.F.-F., A.M.R.,  B.Y., S.A.M., J.P.C.,  J.Y., D.S.J.  All authors discussed results and contributed to the preparation of the manuscript. 
\item[Author Information\!\!\!\!] The authors declare no competing financial interests.  Correspondence and requests for materials should be addressed to K.R.A.H. (kaden.hazzard@colorado.edu).
\end{addendum}
\else
\fi

\ifdefined\Nature
\newpage
\else
\clearpage
\fi

{\center{\Large Supplementary information}}

\ifdefined\Nature
\newcounter{figureS}
\setcounter{figureS}{0}

\makeatletter
\renewenvironment{figure}{\let\caption\NAT@figcaptionS}{}

\newcommand{\NAT@figcaptionS}[2][]{\AtEndDocument{%
    \refstepcounter{figureS}
    \ifthenelse{\value{figureS}=1}{
        \newpage\noindent%
    }{
        \par\vfill
    }
    \sffamily\noindent\textbf{Figure S\arabic{figureS}}\hspace{1em}#2}
    }

\makeatother    

\else
\setcounter{figure}{0}
\renewcommand{\thefigure}{S\arabic{figure}}
\fi

\section{Differential light shift in a 3D lattice}

To enable our use of coherent Ramsey spectroscopy on the sample of ultracold molecules, we create a nearly ``magic'' lattice trap for the two spin states $|${}$\uparrow${}$\rangle$ and $|${}$\downarrow${}$\rangle$, i.e. one that is nearly spin-state-independent~\cite{yan:realizing_2013}. Because the AC polarizability of molecules is anisotropic~\cite{neyenhuis:anisotropic_2012}, the angle between the polarization of the light fields and the quantization axis can be used to tune the light shifts for different molecular states. In Ref.~\cite{yan:realizing_2013}, the details of reducing the differential AC polarizability for the case of $|${}$\uparrow${}$\rangle \equiv |${}$1,-1${}$\rangle$ were reported. The polarizations of the $\hat{X}$, $\hat{Y}$, and $\hat{Z}$ lattice beams were chosen to be linear, at angles of $+45$, $-45$, and $+45$ degrees with respect to the quantization axis, respectively. In this work, we additionally study the case $|${}$\uparrow${}$\rangle \equiv |$$1,0$$\rangle$. In this case, the smallest differential AC polarizability is achieved by setting the linear polarizations to the ``magic'' angle for the states $|$$0,0$$\rangle$ and $|$$1,0$$\rangle$, as determined in Ref.~\cite{neyenhuis:anisotropic_2012}. We find the differential light shift to be minimized when the linear polarizations of the three lattice beams are at angles of roughly $+54$, $-54$, and $+54$ degrees with respect to the quantization axis, respectively.

 \begin{figure}
 \setlength{\unitlength}{1.0in}
\includegraphics[width=3.3in,angle=0]{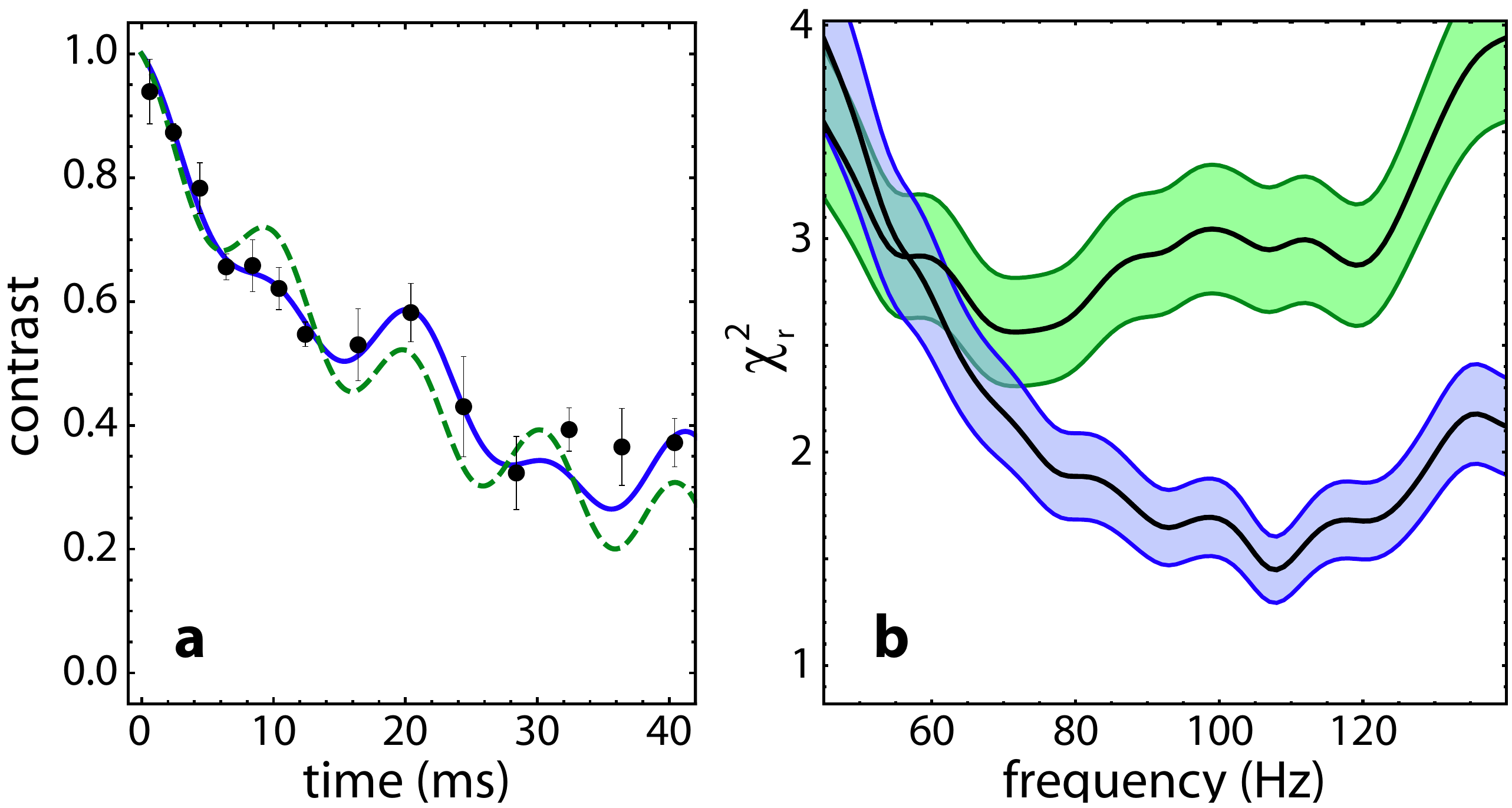}
\caption{\capintro{Fitting oscillations in data.} \labelsubfig{a}~Comparison of the data with empirical fit to contrast oscillations at a single frequency (green dashed line) and three frequencies (blue solid line), as described in the text.
\labelsubfig{b}~Reduced chi-squared ($\chi^2_{\mathrm{r}}$) of multiple data sets for both the single-frequency fitting (green) and three-frequency fitting (blue), as a function of the fit oscillation frequency.
 \label{fig:chisquare}}
\end{figure}

\section{Evidence for multiple interaction energies}

A key feature of the observed spin coherence dynamics is oscillations due to dipole-dipole interactions. While previously 
spin coherence data has been fit to a function with a single oscillation frequency\scite{yan:realizing_2013}, many interaction frequencies are expected due to the long-ranged and anisotropic nature of the dipolar interactions. Here we more comprehensively compare the data to empirical functions that include either a single frequency or multiple frequencies. We simultaneously fit over a dozen data sets that include both sets of spin states, i.e. $\ket{\uparrow}=\ket{1,-1}$ or $\ket{\uparrow}=\ket{1,0}$, rescaling the time axis for the $\ket{1,-1}$ data by a factor of $1/2$ to account for the known difference in dipolar couplings. In considering a single frequency, we fit to the functional form $A \cos^2 (\nu \pi t) + (1-A)\exp (-t/\tau)$, which fixes the contrast to unity at time $t=0$. In considering multiple frequencies, we use our knowledge of the molecules' geometry in the lattice and of the form of the interaction to set the frequency ratios to match those of the 3 strongest pairwise interactions, and fit to the form 
$\sum_i A_i \cos^2(\nu_i \pi t)+(1-\sum_i A_i)e^{-t/\tau}$ with $\{\nu_i\}_{i=1,2,3}=\{\nu,\nu/\sqrt{2},\nu/2\}$.

\begin{figure}
\setlength{\unitlength}{1.0in}
\includegraphics[width=3.35in,angle=0]{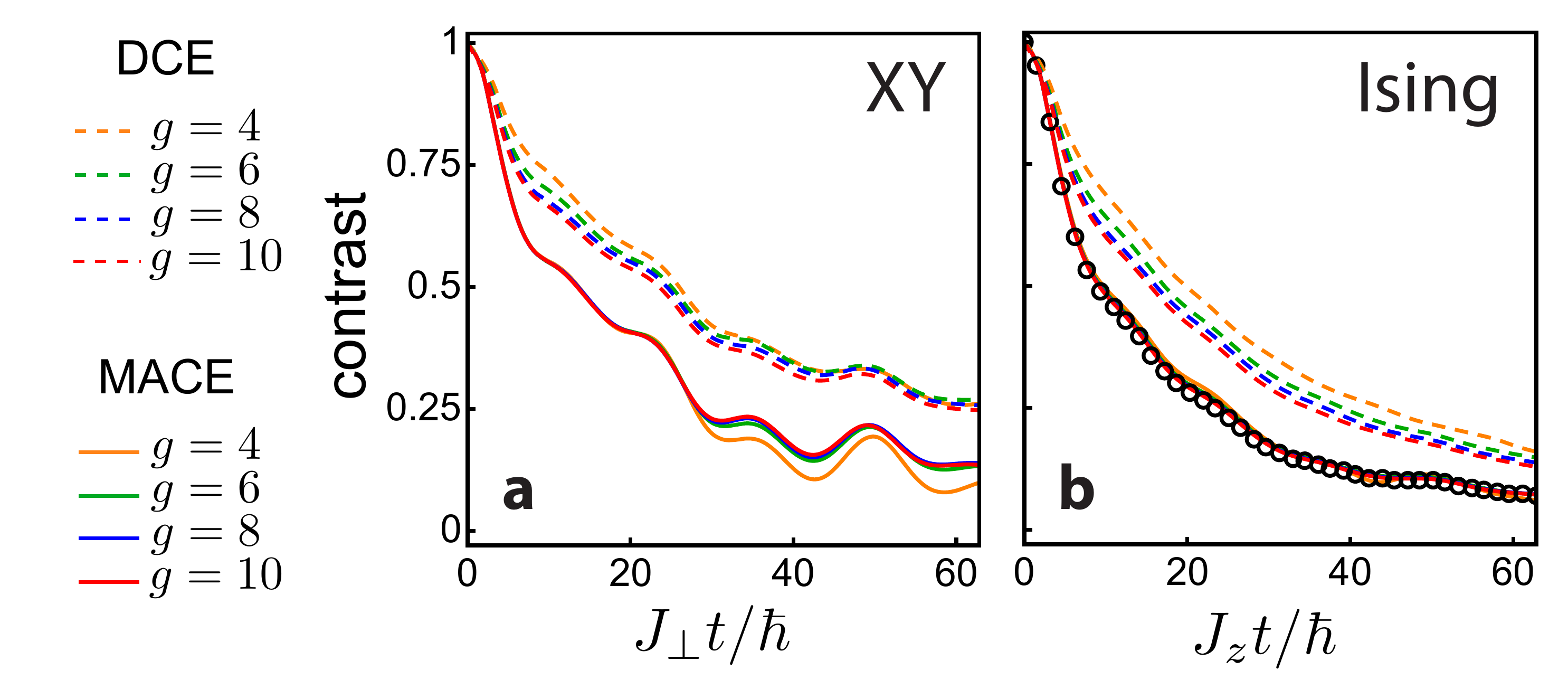}
\caption{\capintro{Convergence of theory and contrast dynamics in XY and Ising spin models.}
Comparison of the MACE [solid lines, schematic in Fig.~\ref{fig:dynamic-protocol}\refsubfig{d}] developed in this paper with the prior state-of-the-art DCE [dashed lines, schematic in Fig.~\ref{fig:dynamic-protocol}\refsubfig{d}], for $N=5,000$ spins and varying cluster size $g$.  The geometry is similar to the experimental one and is described in the text.    Panels~a and~b  show, respectively, XY dynamics [$J_z=0$ in Eq.~\eqref{eq:XXZ-Ham}]
and Ising dynamics ($J_\perp=0$).
For the Ising case, we also show the exact solution (open circles).
 \label{fig:theory-v-theory}}
\end{figure}

In Fig.~\ref{fig:chisquare}~\refsubfig{a}, we show the results of a single-frequency and three-frequency fit for a single data set with $\ket{\uparrow}=\ket{1,0}$. The fit for multiple frequencies clearly does a better job at capturing the observed dynamics, as compared to the single-frequency fit. To make this statement more quantitative, taking into account the additional free parameters allowed for by the multi-frequency fitting, we analyze the cumulative reduced chi-squared ($\chi^2_{r}$) of the fits as applied to all the data sets. The result of this analysis is shown in Fig.~\ref{fig:chisquare}~\refsubfig{b}, as a function of the assumed oscillation frequency. Two major pieces of information can be gleaned from this analysis. First, the fitting with multiple frequencies does a much better job of describing the observed contrast oscillation data, and yields a considerably lower reduced chi-squared value. Second, we find from the frequency dependence of $\chi^2_{\mathrm{r}}$ that the observed oscillations are most consistent with a primary oscillation frequency of $\sim 108$~Hz for the $\ket{1,0}$ data (and by construction a factor of two lower for the $\ket{1,-1}$ case). This value is in good agreement with the theoretical prediction of 104~Hz.

\begin{figure}
\setlength{\unitlength}{1.0in}
\includegraphics[width=3.3in,angle=0]{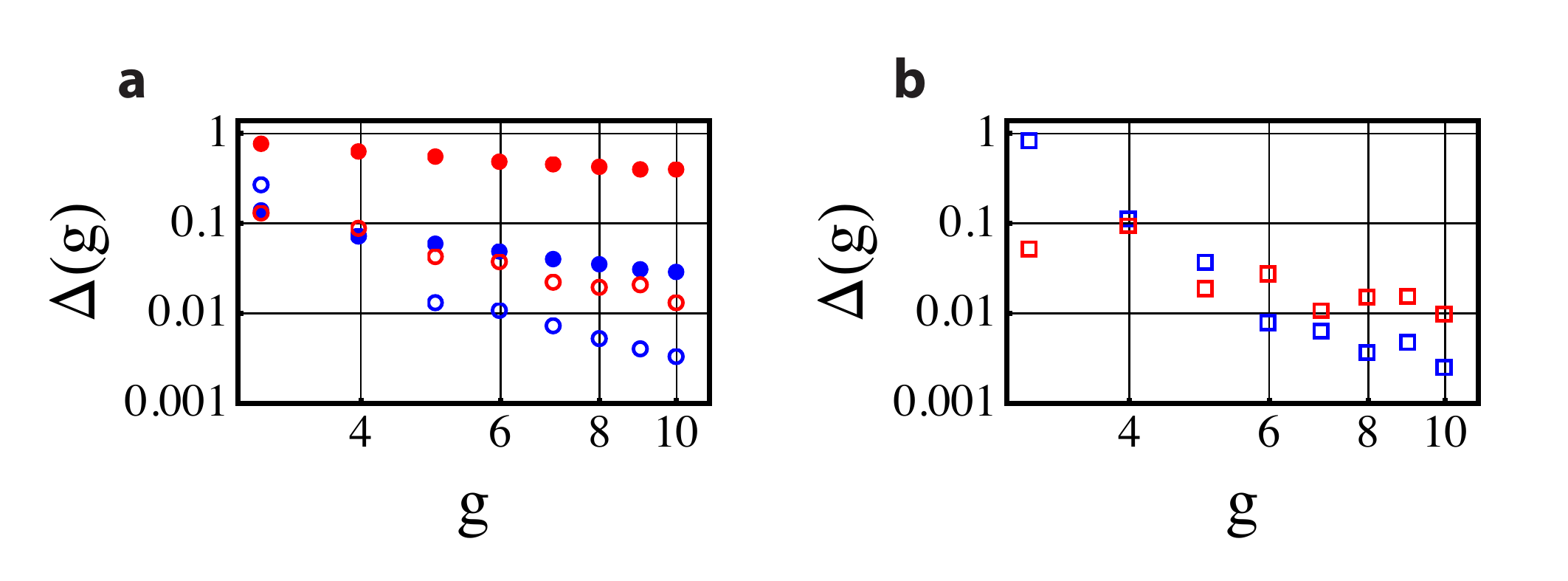}
\caption{\capintro{Theory convergence.}  Root-mean-square (rms) differences of the solutions for cluster sizes $g+1$ and $g$, $\Delta(g)$ (see text), in open symbols (DCE in red and MACE in blue), for \labelsubfig{a} the Ising case and \labelsubfig{b} the XY case.  For the Ising case, we also show the rms difference to the exact solution in filled circles (DCE in red and MACE in blue).  Particle number and the geometry in these calculations are the same as those used for Fig.~\ref{fig:theory-v-theory} in the main text.
 \label{fig:theory-v-theory-rms}}
\end{figure}

\section{Benchmarking theoretical convergence and accuracy}

Figure~\ref{fig:theory-v-theory} demonstrates the superior convergence of the MACE method, and is discussed in detail in the Methods section of the main text. We also study an alternative quantitative measure of the two cluster expansions' convergence: we show (1) for both XY and Ising,  the root-mean-square difference between adjacent cluster sizes and (2) for Ising, the root-mean-square error to the exact solution.  That is, for both cases, we calculate $\Delta(g)=\sqrt{(1/T)\int_0^{T}\! dt\, \lp {\mc C}_{g+1}(t)-{\mc C}_{g}(t)\rp^2}$ where ${\mc C}_g(t)$ is the contrast at time $t$ calculated for cluster size $g$.  For the error to the exact solution, the same formula is used with ${\mc C}_{g+1}$ being replaced by the exact solution.  For the current comparison, we use an integration time $T=30\hbar/J$, with $J=J_\perp$ or $J_z$ as appropriate.  The $\Delta(g)$ versus $g$ are shown in log-log plots in  Fig.~\ref{fig:theory-v-theory-rms}.  Open symbols are adjacent differences, while closed symbols are the difference to the exact theory.  The dramatically lower errors and faster convergence of the MACE are evident.

\end{document}